\documentclass[apj]{emulateapj}
\slugcomment{ApJ Letters Accepted}

\shorttitle{NIR Luminosities in BCGs}
\shortauthors{Batcheldor et al.}

\begin{document}


\title{How Special are Brightest Cluster Galaxies? The Impact of Near-Infrared Luminosities on Scaling Relations for BCGs}


\author{Dan Batcheldor\altaffilmark{1}, Alessandro Marconi\altaffilmark{2}, David Merritt\altaffilmark{3} \& David J. Axon\altaffilmark{3}}
\email{dpbpci@astro.rit.edu}
\altaffiltext{1}{Center for Imaging Science, Rochester Institute of Technology,
    Rochester, NY 14623}
\altaffiltext{2}{INAF-Osservatorio Astronomico Astrofisico di Arcetri, 50125 Firenze, Italy}
\altaffiltext{3}{Department of Physics, Rochester Institute of Technology,
    Rochester, NY 14623}


\begin{abstract}
Using the extended J, H and K magnitudes provided by the 2MASS data archive, we consider the position of brightest cluster galaxies (BCGs) 
in the observed relations between inferred supermassive black hole (SMBH) mass and the host galaxy properties, as well as their position 
in the stellar velocity dispersion and luminosity ($\sigma_{\ast}-L$) relation, compared to E and S0 galaxies. We find that 
SMBH masses ($M_\bullet$) derived from near-infrared (NIR) magnitudes do not exceed $\sim3\times10^{9}M_{\odot}$ and that these masses 
agree well with the predictions made from $\sigma_{\ast}$. In the NIR, there is no evidence that BCGs leave the $\sigma_{\ast}-L$ relation 
defined by less luminous galaxies. The higher SMBH masses predicted from V-band luminosities ($M_\bullet \lesssim 10^{10.5}M_\odot$) are 
attributed to the presence of extended envelopes around the BCGs, however, this will need to be confirmed using deeper multiwavelength imaging. 
\end{abstract}


%% subject header.  Objects should be in the appropriate "individual"
%% headers (e.g. quasars: individual, stars: individual, etc.) with the
%% additional provision that the total number of headers, including each
%% individual object, not exceed six.  The \objectname{} macro, and its
%% alias \object{}, is used to mark each object.  The macro takes the object
%% name as its primary argument.  This name will appear in the paper
%% and serve as the link's anchor in the electronic edition if the name
%% is recognized by the data centers.  The macro also takes an optional
%% argument in parentheses in cases where the data center identification
%% differs from what is to be printed in the paper.

\keywords{galaxies: elliptical --- galaxies: evolution --- galaxies: fundamental parameters --- galaxies: photometry}

\clearpage


\section{Introduction}

Whether the first galaxies were formed from initial large-scale condensations, or grew from an assembly of smaller bodies, still remains 
one of the most fundamental questions in modern astrophysics. Studies of the most massive galaxies will provide important constraints on 
this. Similar considerations apply to supermassive black holes (SMBHs) as the masses of SMBHs correlate with properties of the host bulge 
\citep{fandf05}, i.e., the SMBH mass {\it vs.} bulge luminosity ($M_{\bullet}-L$) relation \citep{kandr95}, the SMBH mass {\it vs.} 
stellar velocity dispersion ($M_{\bullet}-\sigma_{\ast}$) relation \citep{fandm00,geb00}, and the SMBH mass {\it vs.} Sersic index relation 
\citep{gd07}.

As highly luminous massive galaxies found toward the centers of galaxy clusters, brightest cluster galaxies (BCGs) have received 
considerable interest. The surface brightness profiles (SBPs) of BCGs are well fit by the same \citet{ser63} law that describes 
less-luminous spheroids \citep{gra96}, apart from the outer-most regions which sometimes exhibit faint, extended envelopes 
\citep[hereafter B06]{oem76,ber06}. BCGs also appear to obey the same relations between fitting parameters that characterize E/S0 galaxies 
generally \citep{gra96}. \citet[hereafter L06]{lau06a} noted that the $M_\bullet-L$ relation, in the V-band ($M_{\bullet}-L_{\rm V}$), 
predicts higher SMBH masses in BCGs than are predicted by the $M_\bullet-\sigma_{\ast}$ relation. B06 obtain similar results; the slope 
in the size-luminosity relation is found to be steeper in BCGs when compared to the bulk of E/S0's, and the $\sigma_{\ast}$ {\it vs.} 
luminosity ($\sigma_{\ast}-L_{\rm R}$) relation is seen to flatten for the brightest galaxies. 

While the low-scatter $M_{\bullet}-\sigma_{\ast}$ relation is the preferred ``secondary'' SMBH mass estimation technique, compared to the 
larger-scatter $M_{\bullet}-L_{\rm V}$ relation, \citet[hereafter MH03]{mandh03} have shown that the scatter in the relations become similar if 
parameters are derived in the near-infrared (NIR). With this in mind we have conducted a study of BCGs based on the 2MASS extended source 
catalog. We use the 219 L06 galaxies of which $\sim30\%$ are BCGs and the remainder are E/S0s. The L06 data include absolute V-band 
magnitudes ($M_{\rm V}$) and, except in 51 cases, a value for $\sigma_{\ast}$. We adopt the errors of 10\% in $M_{\rm V}$ and $\sigma_{\ast}$, as 
quoted by L06. We supplement the $M_{\rm V}$ data with the NIR data contained within the 2MASS extended source catalog. All magnitudes are 
corrected for galactic extinction according to \citet{sch98}. Distances are all adjusted to a common scale, with 
$H_0=70{\rm~km~s^{-1}~Mpc^{-1}}$, and primarily taken from the survey of \citet{ton01}. Remaining distances are taken from \citet{lai03} 
or from the Virgo in-fall corrected recessional velocities listed by Hyperleda\footnote{\tt http://leda.univ-lyon1.fr} \citep{pat03}. 
In \S~\ref{phot} we evaluate the 2MASS photometry. In \S~\ref{main} we present the results, which are discussed in \S~\ref{dis}. 
\S~\ref{con} sums up.

\section{2MASS Photometry}\label{phot}

\begin{figure}
\plotone{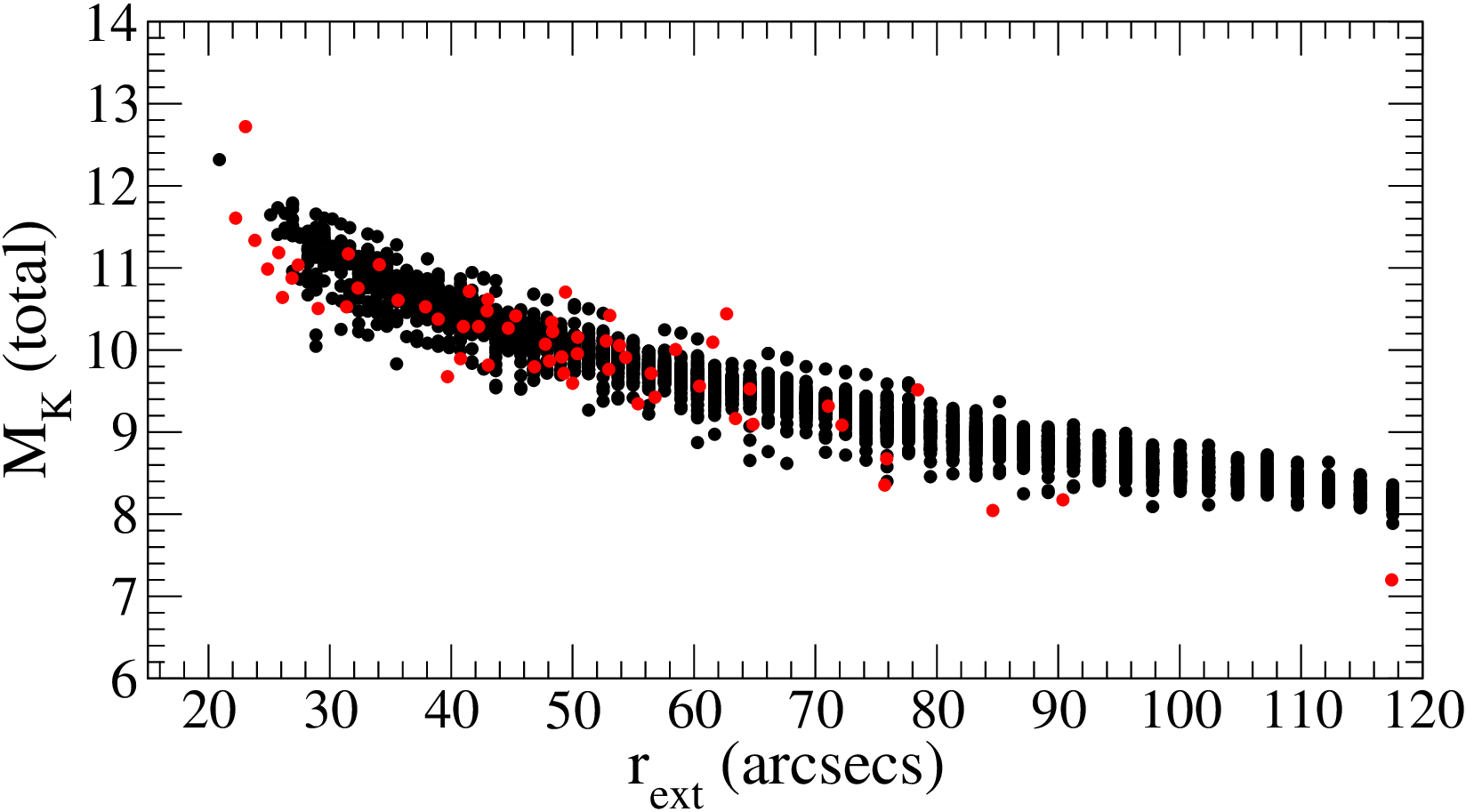}
\caption{Simulations of 2MASS photometry. The distribution of simulated galaxy properties, total K-band magnitude and $r_{\rm ext}$, (black circles) with 
respect to observed BCGs (red circles).}
\label{fig:2mass}
\end{figure}

In this study we have used the 2MASS ``total'' magnitudes (e.g., $k\_m_{\rm ext}$) derived from SBP fitting extrapolation\footnote{For more 
information see sections 2.3a and 4.5e of {\tt http://www.ipac.caltech.edu/2mass/releases/allsky/doc/}}, rather than aperture photometry 
which inevitably under-estimates the total galaxy magnitudes \citep{and02}. Briefly, $k\_m_{\rm ext}$ is estimated by numerical integration 
of the S{\'e}rsic law, fitted between $r>7$--$10''$ (to avoid the point spread function) and the maximum radius ($r_{max}$) of the SBP with 
a signal to noise greater than two. Assuming circular isophotes and $r_{max}=20$--$80''$ (see Figure~\ref{fig:2mass}), this corresponds to a 
SBP limit from 2.9 to 3.7 magnitudes below the 2MASS $3\sigma$ limit (20.09, 19.34 and 18.55 in J, H and K). The best-fitting S{\'e}rsic law 
is then integrated up to $r_{\rm ext}$ (the galaxy ``total'' radius). Typically, $r_{\rm ext}$ is $\sim2$--$5$ times the radius of the 20 
mag/arcsec$^2$ isophote (e.g., $r_{\rm k20}$) where $5\, r_{\rm k20}$ is imposed as a strict upper limit. 

\begin{figure}
\plotone{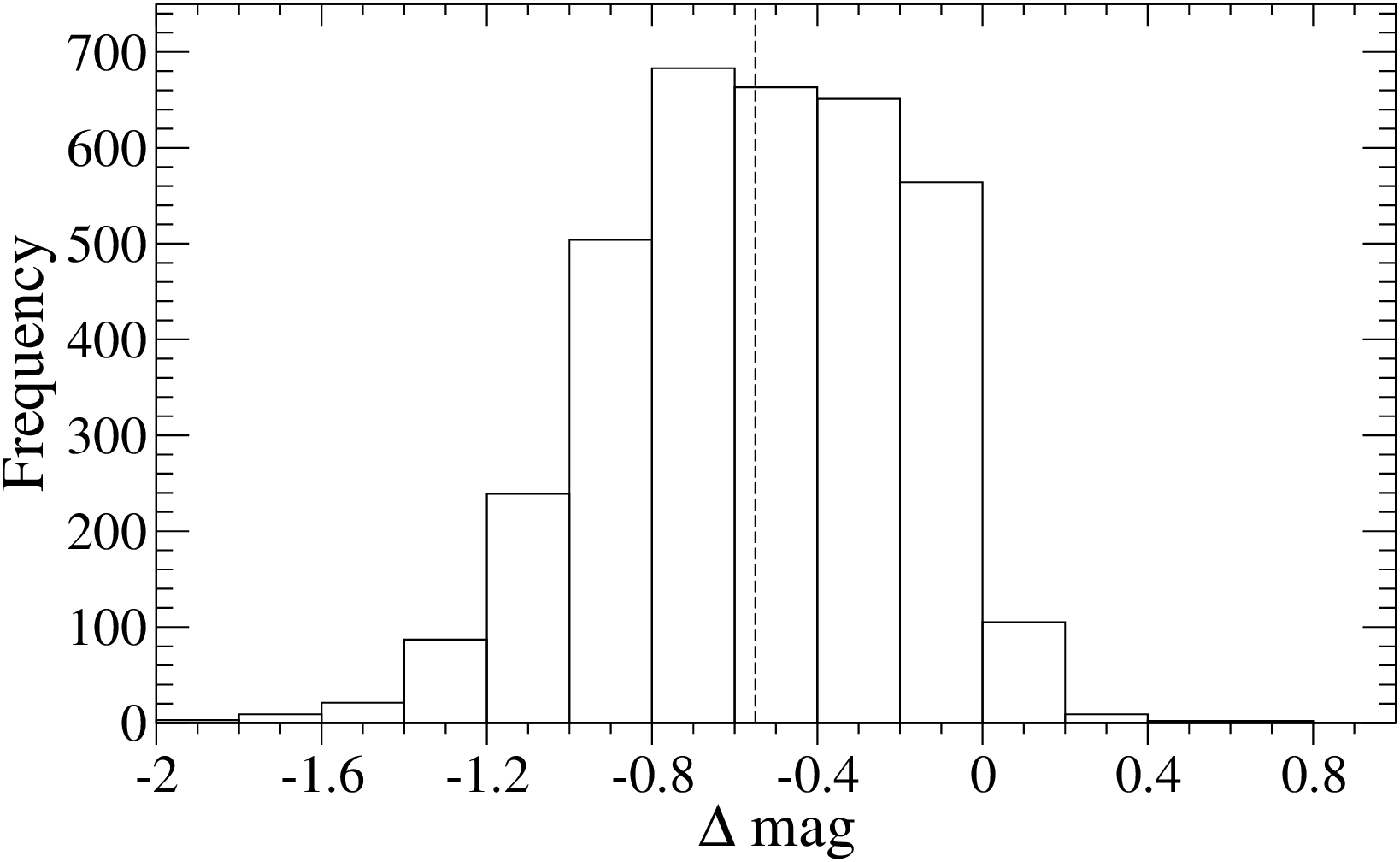}
\caption{Simulations of 2MASS photometry. The difference between ``true'' and estimated magnitudes from $\sim3500$ simulations is shown. The 
dashed line marks the median of the distribution.}\label{fig:2mass2}
\end{figure}                                                        
     
To quantify the importance of undetected light at large radii, we have carried out extensive simulations of BCGs and performed 2MASS analogous 
photometry. Over $3500$ BCGs with $M_{\rm V}$ and $z$ randomly distributed in the observed ranges (-24.5 -- -22 and 0.015 -- 0.050, 
respectively) were generated. For each BCG the effective radius was derived from the $r_{\rm e}-M_{\rm V}$ relation provided by L06, and the S{\'e}rsic 
index, $n$, was derived from the $n-r_{\rm e}$  relation estimated from Figure~11 of \citet{gra96} (0.2 dex scatter was included for both). In 
all cases we imposed $1<n<15$. The observed total K magnitude, $M_{\rm K}$, was then derived using V-K=3.6. The model images of each BCG 
(characterized by $m_{\rm K}$, $r_{\rm e}$ and $n$) were then generated using {\tt GALFIT} \citep{pen02} to take into account the 2MASS pixel sizes 
(1\arcsec) and spatial resolution (FWHM $\sim 3\arcsec$). We added 2MASS typical noise (K = 19.74 mag/arcsec$^2$ rms) and derived SBPs that 
were fitted with a S{\'e}rsic law to estimate $r_{\rm ext}$ and $k\_m_{\rm ext}$. To reproduce the magnitude-size relation (see Figure~\ref{fig:2mass}) 
we used an average $r_{\rm ext}$ of $2.4 r_{k20}$. Figures~\ref{fig:2mass} and \ref{fig:2mass2} show how the simulated magnitudes compare to the 
actual BCG magnitudes and the ``true'' simulated input magnitudes. The average offset is -0.5 mags (50$^{\rm th}$ percentile); 2MASS mildly underestimates 
$M_{\rm K}$ in BCGs. Simulations using bluer colors (V-K=3.3) give very similar result.

\begin{figure}
\plotone{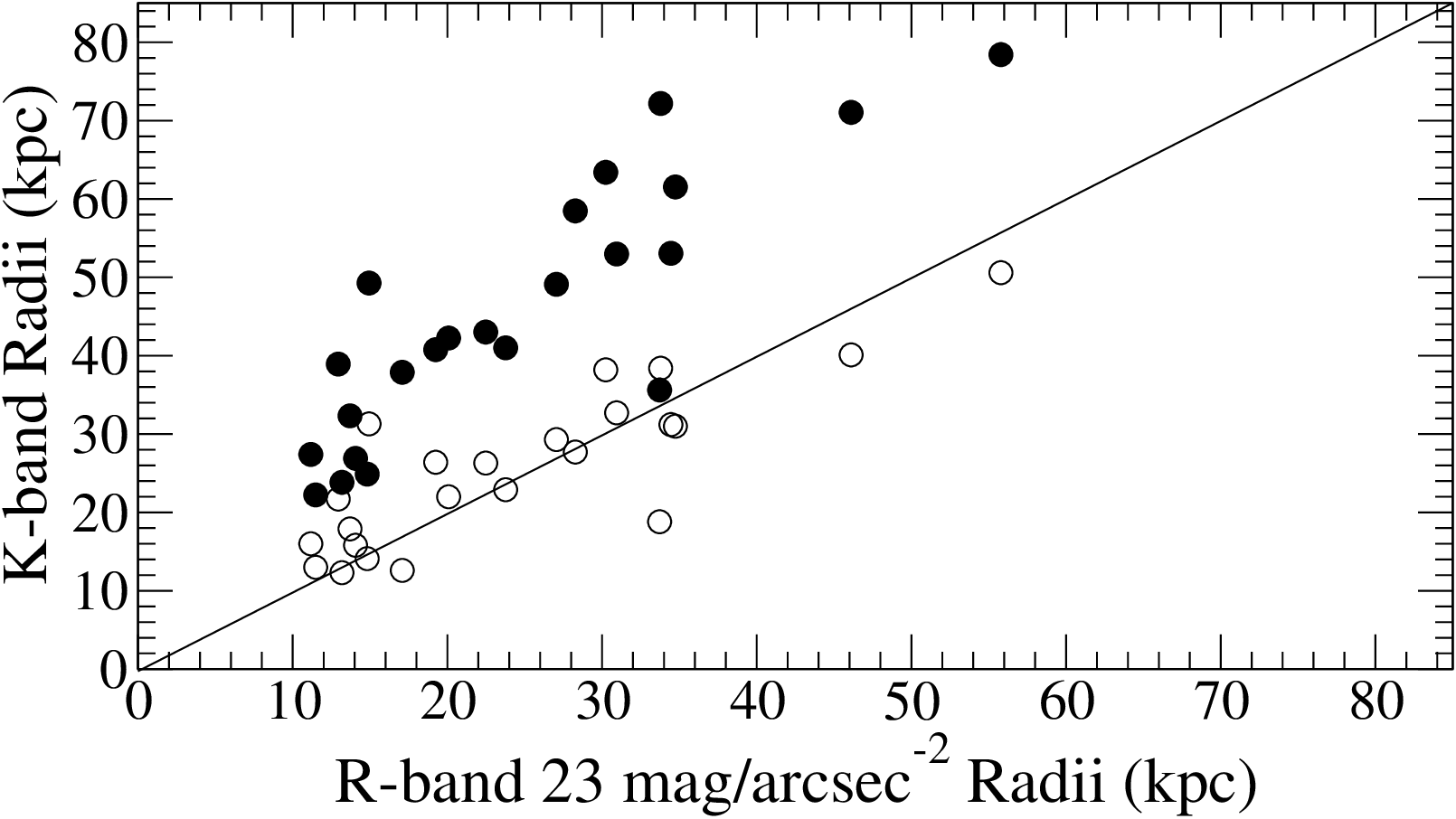}
\caption{Comparing the radii of V-band and NIR magnitudes. The solid line is one-to-one. Open and closed circles show the 
relation between $r_{\rm r23}$ and $r_{\rm k20}$ and $r_{\rm ext}$ respectively.\label{fig:comprad}}
\end{figure}
 
\begin{figure}
\plotone{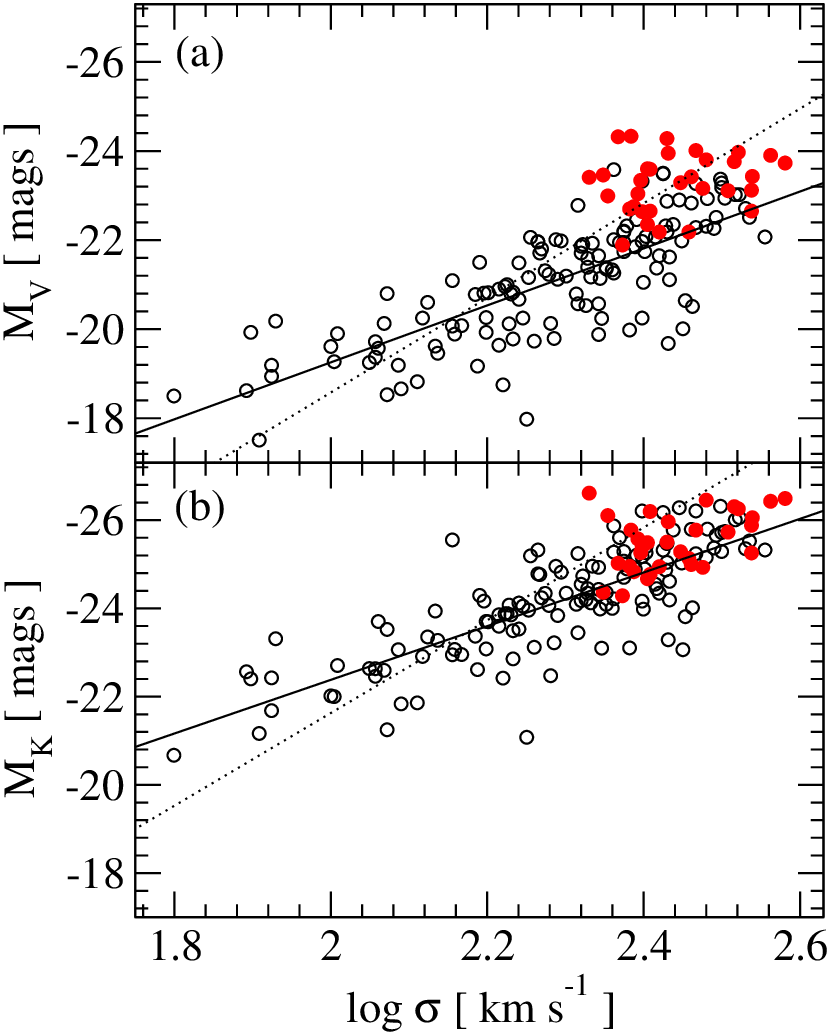}
\caption{The relationship between $\sigma_{\ast}$ and luminosity in the V-band (a) and K-band (b). Open black circles mark E and S0s, 
filled red circles show BCGs. The solid lines show the fit to just the E/S0s, the dotted lines mark the fit of \citet{ber06}.}
\label{fig:lsig}
\end{figure}
 
In \S~\ref{main} we will directly compare the results from extrapolating photometry estimated in the V-band and NIR. As the V-band 
photometry may have been gained from different radii (thereby gathering a different amount of light) we now briefly compare the relative 
sizes of the regions from which the photometry was derived. The L06 V-band BCG luminosities were provided by \citet{lai03}, who in turn 
estimated the total magnitudes from SBPs presented by \citet{pl95}. \citet{gra96} performed S{\'e}rsic fits to these SBPs using a limiting 
R-band surface brightness. Therefore, assuming a color of R-K=3.0, we can directly compare the radii of $r_{\rm ext}$ and $r_{k20}$ (the radius of 
the K=20 mag/arcsec$^2$ isophote) with the R-band surface brightness at 23.0 mag/arcsec$^2$ ($r_{\rm r23}$). By calculating $r_{\rm r23}$ from 
the \citet{gra96} BCG S{\'e}rsic fits, we find that $r_{\rm k20}$ agrees well with the $r_{\rm r23}$ radii (Figure~\ref{fig:comprad}). However, 
$r_{\rm ext}$ is always larger than $r_{\rm r23}$. Therefore, the 2MASS magnitudes used in this study include light {\it from at least} as extended 
a region as the V-band magnitudes used by L06.
     
As an additional check, we can compare published values of $M_{\rm K}$ to those derived by MH03 from two-dimensional fitting to 2MASS 
profiles out to infinity. We find a systematic offset of $\sim0.4$ mags, consistent with our simulations. 

\begin{figure}
\plotone{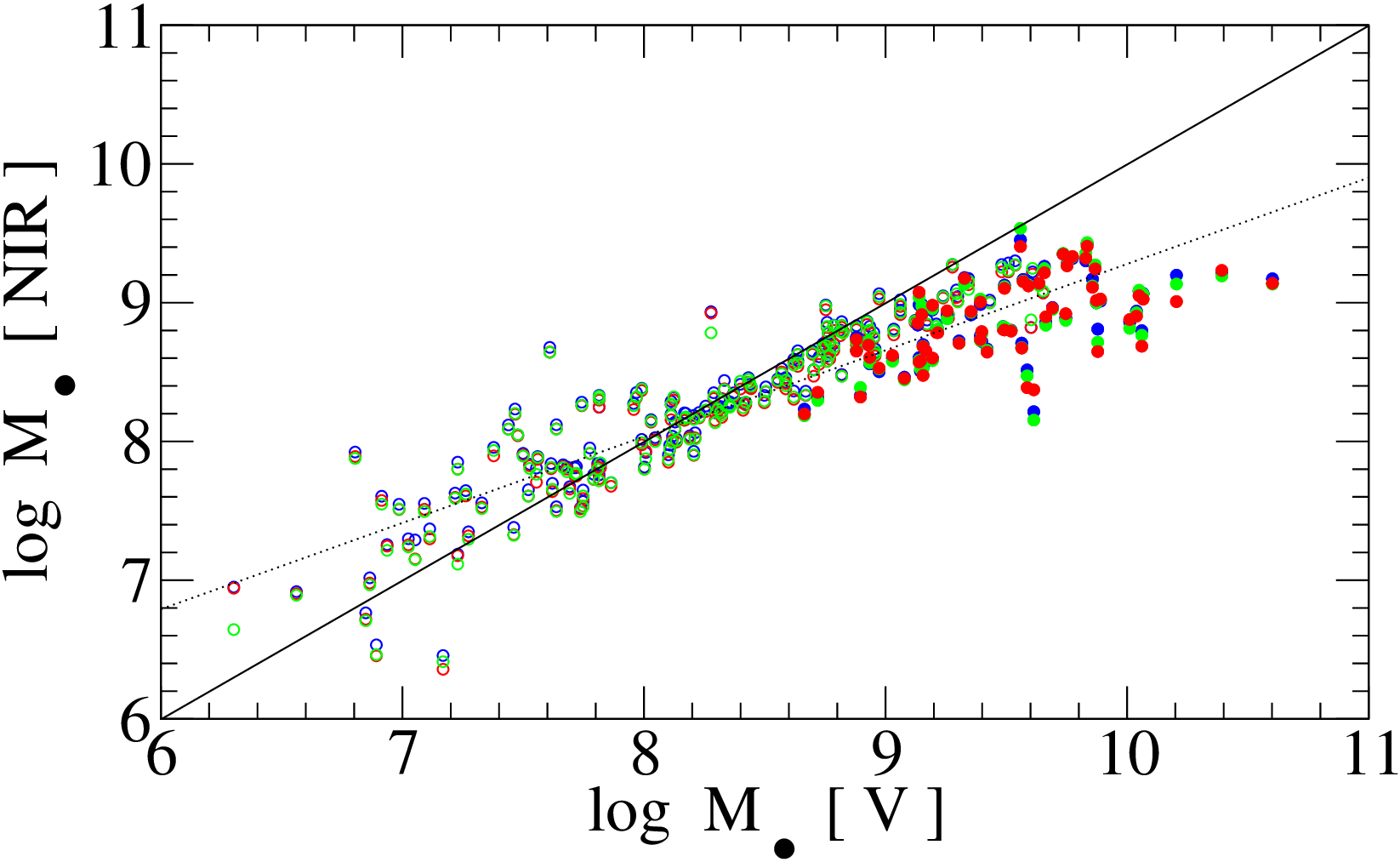}
\caption{V-band {\it vs.} NIR $M_{\bullet}$ estimates. The solid line marks a one-to-one relation. Open circles are E/S0's, closed 
circles are BCGs. Blue, green and red colors refer to the J, H and K bands respectively. The dotted line marks the fit to all the 
K-band data.\label{fig:vvsnir}}
\end{figure}

\section{BCGs in the NIR}\label{main}

Figure~\ref{fig:lsig} compares the $\sigma_{\ast}-L_{\rm V}$ and $\sigma_{\ast}-L_{\rm K}$ relations. In both cases we plot the 
best-fit relation defined by the E/S0 galaxies (solid line) as well as the $\sigma_{\ast}-L_{\rm R}$ fit given by B06 (their Figure~6 with 
colors of V-R=0.6 and R-K=3.0). We find shallower slopes for the E/S0 population consistent with B06. Figure~\ref{fig:lsig}(a) 
demonstrates that the ``bending'' of the $\sigma_{\ast}-L_{\rm V}$ relation, as noted by B06, is also seen in the L06 sample; BCGs fall above 
the relation defined by the E/S0s. In the NIR (Figure~\ref{fig:lsig}b) the BCGs do not appear to define a separate population; instead their 
distribution is indistinguishable from that of the E/S0s. The average offset of BCGs from the E/S0 relation is 1.20 mags in $M_{\rm V}$ and 0.48 mags 
in $M_{\rm K}$.

Figure~\ref{fig:vvsnir} presents the relationship between $M_{\bullet}$ estimated from the V-band (hereafter $M_{\bullet}(V)$), using the 
relation as defined by L06 (their Equation~4), and $M_{\bullet}$ estimated from the NIR data (hereafter $M_{\bullet}(J,H,K)$) using the 
relations defined by the MH03 sample. The upper limit for $M_{\bullet}(V)$ is $10^{10.5}M_{\odot}$. Below $M_\bullet\approx10^{8.5}M_{\odot}$ 
the agreement between all bands is good. Above $10^{8.5}M_{\odot}$, the NIR data predict significantly lower SMBH masses, with none exceeding 
$10^{9.4}M_{\odot}$. The fit to the K-band relation is shown as a dotted line and has a slope of $0.62\pm0.02$ ($0.74\pm0.02$ for E/S0s). For 
estimates of $M_{\bullet}(J,H,K)$ we do not use the exact fits presented by MH03 (e.g. their Table 2) as they were derived using a bisector method
(e.g., assuming scatter in both $M_{\bullet}$ and $L_{\rm J,H,K}$). Instead, $M_{\bullet}(J,H,K)$ has been determined from a single Y$|$X fit to 
$M_{\bullet}$ and $L$ (taking into account errors from both axis) because in estimating $M_{\bullet}$ from $L$ one assumes that all the scatter is 
in only one variable. The form of this relation, in the K-band, is given by $\log{M_{\bullet}}=8.22\pm0.07+(1.06\pm0.11)(\log{L_K}-10.9)$. The results, 
in this case, are insignificant from the original MH03 fits ($\sigma=10^{0.04}M_{\odot}$).

Figure~\ref{fig:dirsigmag} shows how photometric $M_{\bullet}$ estimates compare to those from the $M_{\bullet}-\sigma_{\ast}$ 
relation. The \citet{tre02} expression is used to derive $M_{\bullet}$ from $\sigma_{\ast}$, hereafter $M_{\bullet}(\sigma_{\ast})$. 
In Figure~\ref{fig:dirsigmag}(a) the $M_{\bullet}-L_{\rm V}$ relation (above $10^{8.5}M_{\odot}$) predicts SMBH masses greater than those expected 
from the $M_{\bullet}-\sigma_{\ast}$ relation. However, in the NIR, this observation is not made; both predictions are consistent. The 
$M_{\bullet}-L_V$ relation implies $M_{\bullet}\lesssim2.5\times10^{10}M_{\odot}$, whereas the NIR produces 
$M_{\bullet}\lesssim2.8\times10^{9}M_{\odot}$. The scatter in the $M_{\bullet}(J,H,K)-M_{\bullet}(\sigma_{\ast})$ relations are significantly less 
than the $M_{\bullet}(V)-M_{\bullet}(\sigma_{\ast})$ relation. 

The effect of the results from the \S~\ref{phot} simulations can be seen by artificially and randomly introducing the distribution 
of $\Delta$ mag to the K-band magnitudes. A comparison of $M_{\bullet}(\sigma_{\ast})$ and the adjusted $M_{\bullet}(K)$ is shown in 
Figure~\ref{fig:simmag}. The slope of the best fit relation (dotted line) is $0.85\pm0.05$ and no values of $M_{\bullet}(K)$ exceed $10^{9.7}M_{\odot}$.

\section{Discussion}\label{dis}

\begin{figure*}
\plotone{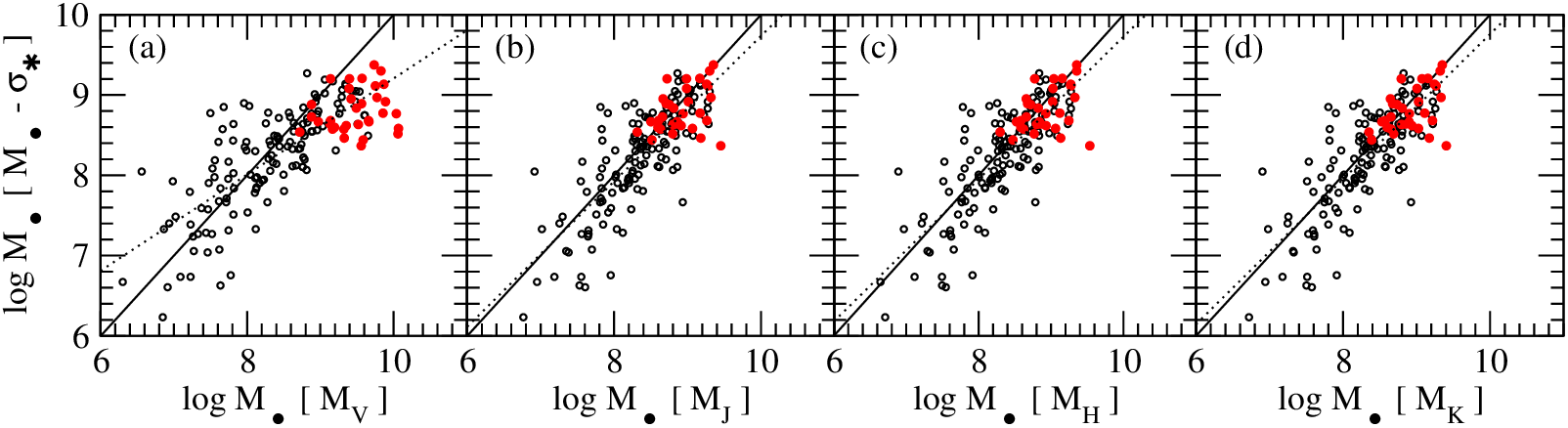}
\caption{Comparing photometric $M_{\bullet}$ estimates with $M_{\bullet}-\sigma_{\ast}$ estimates. V, J, H and K masses (a, b, c and d 
respectively) are compared to $M_{\bullet}-\sigma_{\ast}$ estimates. Open black circles are E/S0s, BCGs are filled red circles. In all cases 
the solid line represents a one-to-one relation and the dotted line the best-fit relation.\label{fig:dirsigmag}}
\end{figure*}

\begin{figure}
\plotone{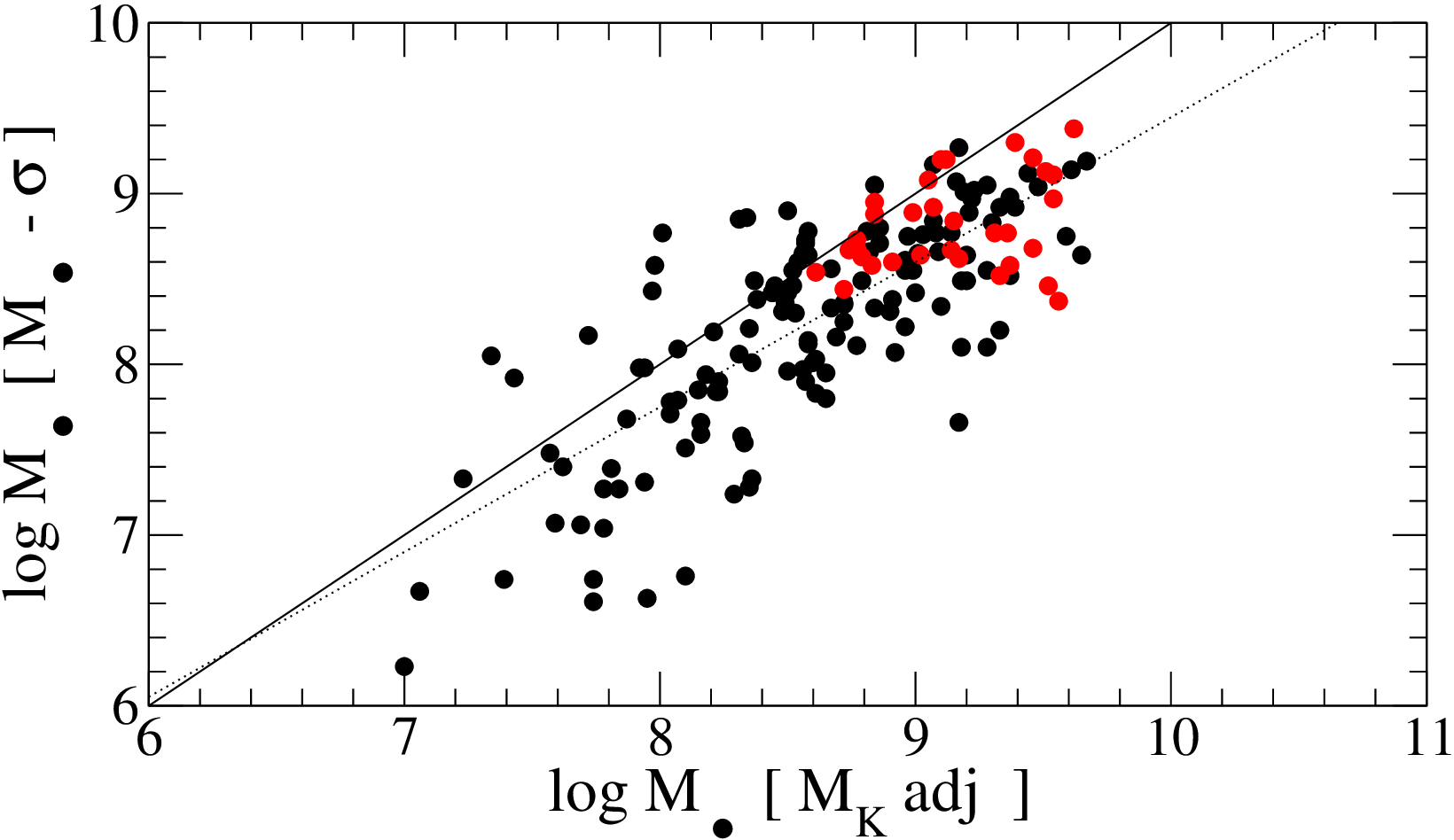}
\caption{Comparing redistributed  $M_{\bullet}(K)$ estimates with $M_{\bullet}-\sigma_{\ast}$ estimates. Black circles are E/S0s, BCGs are red circles. The 
solid line represents a one-to-one relation and the dotted line the best fit relation.\label{fig:simmag}}
\end{figure}

It is evident from both Figures~\ref{fig:lsig} and \ref{fig:dirsigmag} that the dispersion of BCGs in the NIR is considerably less than 
in the V-band. This is also the case for the E/S0s. Even if we are underestimating the NIR luminosities, it would require a very fortunate 
coincidence to have such a strong agreement between $M_\bullet(\sigma_{\ast})$ and $M_{\bullet}(J,H,K)$ across the entire mass function. 
It then follows that BCGs are not ``special'' when viewed at NIR wavelengths. BCGs follow the same $\sigma_{\ast}-L_{\rm K}$ distribution as 
defined by less luminous spheroids, and comparable masses are predicted for the SMBHs in BCGs based either on velocity dispersions or on 
total magnitudes.

We have demonstrated, through extensive simulations, that the 2MASS magnitudes used in this study are robust to within 0.5 mags for BCGs. 
This is consistent with the offset between the 2MASS photometry and that of MH03, who take the total magnitude components of S{\'e}rsic fits 
to two-dimensional photometric models. This offset is expected as the 2MASS total magnitude integrations stop at $r_{\rm ext}$. For the 
$M_{\bullet}-L_{\rm J,H,K}$ relations to predict a population of $10^{10}M_{\odot}$ SMBHs, i.e., for the BCGs to fall on the one-to-one 
relation in Figure~\ref{fig:vvsnir}, 2MASS photometry would have to be increased, on average, by 1.65 mags. This offset is inconsistent 
with the 2MASS underestimates. The offset would also need to vary with $M_{\bullet}$ and have no effect at masses below 
$10^{8.5}M_{\odot}$ where the results between $M_{\bullet}-L_{\rm V}$ and $M_{\bullet}-L_{\rm J,H,K}$ are consistent. Conversely, the V-band 
photometry must be overestimated by 2.17 mags, on average, for the upper limit of $M_{\bullet}$ to be similar to that predicted from 
the NIR. 

Why then do BCGs show an excess of V-band light over their less massive cousins? The issue could be resolved by considering the faint 
extended luminous halos known to surround the most massive galaxies \citep{oem76}. The signature of such halos is an inflection in the 
SBPs at large $r$. In the NIR, the total magnitudes could be missing the contributions from faint halos or, alternatively, V-band 
luminosities could be interpreted as over-estimates due to the spurious inclusion of halo light. Since extended halos predominantly 
sit in the overall cluster potential, they are unlikely to be related dynamically to the central regions from which the BCG 
$\sigma_{\ast}$ is typically measured. Photometry from further toward the blue end of the spectrum may be deep enough to include 
a significant contribution from these extended halos, leading to an increase in the estimation of BCGs luminosities and a turn-over in 
$M_{\bullet}$ estimates above a certain threshold. 

While we have shown that BCGs are not ``special'' in terms of their $\sigma_{\ast}-L_{\rm K}$ distribution, nevertheless, their extreme 
luminosities and unique locations close to the centers of galaxy clusters suggest special formation processes. It has long been 
argued that the extended envelopes of BCGs are debris from tidally-stripped galaxies, and hence that they are associated more closely with 
the overall cluster potential well than with any single galaxy \citep{mer84a}.  The envelopes may also consist in part of stars formed in 
cooling flows \citep{fab94}. The presence of multiple nuclei in some BCGs argues in favor of these galaxies not being fully relaxed 
\citep{mer84b}. A photometrically complete, high-resolution imaging survey of BCGs would be able to provide a framework for a more 
quantitative analysis of these fundamentally important objects.

\section{Conclusions}\label{con}

Brightest cluster galaxies offer the chance to study the pinnacle of galaxy evolution. They also give us the opportunity to study the 
top of the SMBH food chain by using the observed relations between $M_{\bullet}$ and the properties of the surrounding host galaxy. 
We have shown that NIR luminosities, combined with previously established scaling relations (MH03), imply a maximum mass of 
$\sim3\times10^{9}M_{\odot}$. This is consistent with the most massive SMBH directly modeled at the center of M87 \citep{mac97} and with 
the direct $M_{\bullet}$ estimates of \citet{dal06} in 3 BCGs. We also find that, across all values, SMBH masses predicted using NIR magnitudes 
are consistent with masses predicted from $\sigma_{\ast}$. In addition, we have shown that BCGs follow the same distribution, as defined by E/S0 
galaxies, in the $\sigma_{\ast}-L_{\rm K}$ relation. If confirmed by a deep multiwavelength study, these findings could have important implications 
for the nature of the SMBH mass function, and, as in the past, would show that NIR data are to be preferred when estimating $M_{\bullet}$ (MH03). 
While BCGs are likely not special -- in the sense of hosting hyper-massive black holes or by defining a distinct population in the 
$\sigma_{\ast}-L_{\rm K}$ plane -- they may be interesting by virtue of being surrounded by extended faint halos. The unique local BCG 
environment, deep within a cluster potential, could be the generator of these halos, which may be populated by younger stars tidally stripped 
from other cluster members or that are the results of intra-cluster gas accretion or other recent merger events. 

\acknowledgments

We would like to thank Alister Graham, Tod R. Lauer and Mariangela Bernardi for their comments on this manuscript. This research used 
NASA/IPAC Infrared Science Archive, which is operated by the JPL, CalTech, under contract with the NASA. We acknowledge the usage of the 
HyperLeda database (http://leda.univ-lyon1.fr). D.M acknowledges support from grants AST 00-71099, AST 02-06031, AST 04-20920, and 
AST 04-37519 from the NSF, grant NNG04GJ48G from NASA, and grant HST-AR-09519.01-A from STScI.

\end{document}